The reliability of an environmental epidemiology meta-analysis, a case study
S. Stanley Young, Mithun Kumar Acharjee, Kumer Das

https://doi.org/10.1016/j.yrtph.2018.12.013

**Summary**

**Background** Claims made in science papers are coming under increased scrutiny with many claims failing to replicate. Meta-analysis studies that use unreliable observational studies should be in question. We examine the reliability of the base studies used in an air quality/heart attack meta-analysis and the resulting meta-analysis.

**Methods** A meta-analysis study that includes 14 observational air quality/heart attack studies is examined for its statistical reliability. We use simple counting to evaluate the reliability of the base papers and a p-value plot of the p-values from the base studies to examine study heterogeneity.

**Findings** We find that the based papers have massive multiple testing and multiple modeling with no statistical adjustments. Statistics coming from the base papers are not guaranteed to be unbiased, a requirement for a valid meta-analysis. There is study heterogeneity for the base papers with strong evidence for so called p-hacking.

**Interpretation** We make two observations: there are many claims at issue in each of the 14 base studies so uncorrected multiple testing is a serious issue. We find the base papers and the resulting meta-analysis are unreliable.

**Funding** The work of Young was partially funded by the American Petroleum Institute. The other researchers are unfunded.

**Introduction**

A positive result for a single, sharp question that is examined in a randomized experiment is usually taken as evidence for causality. Replication increases confidence. Randomized studies are either not feasible or have not been done for many important questions, medical and environmental. Where randomized studies are not feasible, researchers often combine results from observational studies into a meta-analysis. It is thought that a meta-analysis will have improve reliability over and above what is available in any of the individual studies. In practice, is that true? Advice on how to conduct and report on a meta-analysis of observational studies is given by Stroup et al. (2000). These authors are largely silent assessing the reliability of the individual observational studies.

Claims coming from many science disciplines fail to replicate, Horton (2015) among many others. The failure of observational studies to replicate is a well-known, old problem, Mayes et al. (1988). They looked at 56 questions that were examined by the case-control method and the papers divided roughly equally on each question. Clearly, half of the studies were wrong. Unlimited slicing and dicing of a data set as a problem has been long recognized, Cournot (1843). Feinstein (1988) restated the question and noted that not defining and not limiting the questions in a study was likely part of the problem. Rothman (1990) quickly asserted that multiple testing was not an issue, and many continue to subscribe to his position. More recently Perneger (1998) asserted "the view, widely held by epidemiologists, that Bonferroni adjustments are, at best, unnecessary and, at worst, deleterious to sound statistical inference." Both Rothman and Perneger are highly cited by authors that want to leave no stone unturned and, to that end, are happy to have any number of false positives, a position that is extremely self-serving for the producer and a disaster for the consumer. For a criticism of Rothman/Perneger, see Frane (in press).

Using a statistical argument, Ioannidis (2005) asserted that 90% of sciences claims were expected not to replicate. Young and Karr (2011) found 12 randomized clinical studies where 52 claims coming from observational studies were tested. None of the 52 claims were statistically significant in the claimed direction while five were significant, but in the opposite direction. Begeley and Ellis (2012) noted that 47 of 53 claims coming from experimental biology failed to usefully replicate. A Nature survey of ~1,500 scientists reported that 90% thought there was a crisis, 52% serious, 38% minor, Baker (2016).

Three recent books, Hubbard (2015), Harris (2017) and Chambers (2017), henceforth referred to as HHC, discuss statistical, experimental and logical problems with current science. Multiple testing is one of the problems. If you ask a lot of questions and base your decisions on a statistically test at the 5% level, false positives are expected. Pejoratively, asking a lot of questions is called p-hacking. Head et al. (2015): "We used text-mining to search for p-values in all Open Access papers available in the PubMed." and concluded that p-hacking is rife. Naming your hypothesis after the fact, letting the data suggest a hypothesis for any found result, is called, HARKing, Hypothesis After the Results are Known. The problems p-hacking and HARKing were noted, but not so named by Feinstein (1988).

Standard science is to propose a single, sharp hypothesis before the fact and then collect data, either experimental or observational, to test the sharp hypothesis. Randomized clinical trials, as supervised by the Food and Drug Administration, are a good example. Karl Popper is one of the great philosophers of science. His position was that a theory could not be proven correct, but could be tested by sharp, decisive experiments and proven wrong. HARKing is a total corruption of the (Popper) science process.

Some assert that the p-value itself is the problem and that hypothesis testing needs to be abandoned, Wasserstein RL, Lazar NA. (2016) or that the significance level used needs to be dramatically reduced, Boos and Stefanski (2011) Johnson (2013), 0.001 and 0.005 respectively. These authors assume one, pre-planned question with careful attention to study design and study execution. Hubbard 2015 would move away from or abandon hypothesis testing altogether and concentrate on replication of findings as the path to reliable results. Any sharp decision rule should take multiple testing into account. We assert that a major problem is multiple testing.

In this paper, we examine a meta-analysis study, Nawrot et al. (2011), hereafter Nawrot, where their focus is on the question: can Particulate Matter, PM, induce a heart attack? PM is ubiquitous and is an air component regulated by the US Environmental Protection Agency. PM is not chemically defined; physically, it consists of particles small enough to pass through a 10 micron grid. Note that Nawrot assumes that PM10 is 70% PM2.5 (See their Table 3.), so their analysis applies to both. Many observational studies show an association between PM and heart attacks (myocardial infarcts, Mis); 14 of these papers were collected by Nawrot.

First, we examine the reliability of the 14 observational studies that deal with PM and that are used in the Nawrot meta-analysis study. We use simple counting to establish the size of the analysis search space available to the researchers. We look at the distribution of the results from these papers, p-values, to evaluate the possibility of p-hacking/p-HACKing, building on ideas of Simonsohn et al. 2014.

**Methods**

Our methods involve simple counting for each base study to examine the analysis search space and the examination of the distribution of p-values across the studies using a p-value plot. We count the numbers of outcomes, predictors, covariates and lags and use those counts to approximate the analysis search space available to a researcher, the number of possible statistical tests. Many researchers think that air quality some days before the day at issue might give rise to a heart attack, hence the interest in lags. Mostly, this counting is easy, but it can take some time to count these features as the authors might mention some of the variables anywhere in the paper. Covariates are often given in different places and some available covariates might not be mentioned at all in the paper.

The product of Outcomes, Predictors and Lags gives the number of questions at issue, Space_1. A covariate can be in the model or not so one way to approximate the modeling options is to raise 2 to the power of the number of covariates, Space_2. The product of Space_1 and Space_2 gives an approximation to analysis total search space, Space_3.

The meta-analysis researcher takes a summary statistic from each paper to combine in the meta-analysis estimation of the effect in question. In this instance, the summary statistic is a risk ratio with confidence limits. For 12/14 studies, unadjusted p-values are reported; two studies note their results were not significant. Each base paper tested at the 0.05 level, ignoring multiple testing. The authors of Nawrot take the summary statistics and their p-values at face value. The 14 risk ratios are taken from Nawrot, their Table 3 and their p-values are computed using their

95% confidence intervals. If there is no effect of PM10 on heart attacks, these p-values should follow a uniform distribution. The distribution of the p-values can be examined using a p-value plot, Schweder and Spjøtvoll (1982). In a p-value plot, the p-values are rank ordered from smallest to largest and plotted against the integers, 1, 2, 3, …, 14. If the p-values form a 45-degree line, then that is evidence for randomness, no effect. The positioning of p-values in the distribution can be used to evaluate potential manipulation of the analysis process, Simonsohn et al. (2014); they worked with a limited the range of p-values, those less than or equal to 0.05 as those were the only ones reported in their literature. We have p-values across the whole interval, so we use them all in a p-value plot.

**Results**

Table 1 gives the counts of the outcomes, predictors, lags and covariates for the 14 papers. Table 2 gives summary statistics for Space_1, Space_2 and Space_3. The median of Space_3, the number of possible analyses, is 6,784 with an interquartile range of 2,600 to 94,208. The modeling search space can be large. The authors of the 14 papers screened statistically using a p-value of 0.05; the authors provide only very weak evidence to support their claims. Note that a small p-value due to chance will have an associated biased statistic.

Table 3 gives the named covariates in the papers used by Nawrot. Some covariates are used a lot, e.g. age and sex, and some version of temperature/relative humidity, while others are particular to a paper. There is considerable variation in how weather variables are used.

A p-value plot of the 14 p-values is given in Figure 1. We see many small p-values, but we also see p-values that roughly form a 45-degree line that are consistent with no effect. A test for a quadratic dose response gave a p-value of 0.00015, which supports a bi-linear response. The p-values are heterogeneous; some are nominally statistically significant, while others appear to be random.

**Discussion**

A basic requirement of a meta-analysis that a statistic, mean value, risk ratio, etc., taken from a base paper and use in a meta-analysis be an unbiased estimate of the statistic of interest, Boos and Stefanski (2013). In the Nawrot meta-analysis, the authors took the estimates from the base papers at face value, which assumes that they are unbiased, and which we think is unsupported given the size of the search spaces in the base papers. The authors of the base papers have the responsibility to provide solid statistical evidence to support their statistically based claims. In all the base papers, statistical testing was at the 0.05 level while large numbers of claims were at issue. If the small p-value reported in the base paper is due to multiple testing/multiple modeling, then the associated statistic biased. The authors of the base papers provide no evidence that the small effects they report are anything but chance. In the 14 papers used by Nawrot to examine the relationship of PM10 to heart attacks, there was no adjustment for multiple testing.

The distribution of the p-values from the base studies was evaluated using a p-value plot. If you take the p-values from each of the base questions and look at the distribution of them, you can evaluate the possibility of analysis manipulation. These observational studies are large and complex and the researcher can try alternative ways to do the analysis; some of these analysis might give a p-value less than 0.05, opening the door to publication.

If when looking at multiple p-values, one from each analysis, you find many below the publishable 0.05, but the remaining p-values appear uniformly distributed between 0.05 and 1.00, then you have a problem. Do you believe the small p-values or the not significant p-values?

IF there is a real effect, the non-significant ones should still be relatively small and still fall on a line with a slope less than 45 degrees. If p-values are bi-linear and the analysis search space is large, then there is evidence for analysis manipulation in the base papers.

Editors and referees generally require a statistically significant result to support publication, HHC. Simonsohn et al. 2014 support this statement but go on to say that authors might search through many results until they find a significant p-value, p-HACKing. Also, authors might simply not submit a negative study, the file drawer problem, leading to publication bias.

Is there additional support for a negative finding? Milojevic et al. (2014) looked at cardiovascular disease, CVD, events in all of England and Wales. "…over 2 million CVD emergency hospital admissions and over 600 000 CVD deaths" were linked to six air quality variables including PM2.5 and PM10 and tested for 11 cardiac diagnoses. Hospital admissions and mortality were analyzed. There were 6x11x2=132 questions at issue. Their Figures 1&2 are given here as our Figures 2&3. Percent change in hospital admissions and mortality are the y-axes in the glyphs. Along the x-axis are given the standard air components, carbon monoxide (CO), nitrogen dioxide (NO2), particulate matter less than 10 μm in aerodynamic diameter (PM10), particulate matter less than 2.5 μm in aerodynamic diameter (PM10) and sulfur dioxide (SO2 ), and daily maximum of 8-hourly running mean of O3 measured at the nearest air pollution monitoring site to the place of residence. The results given for PM10 are consistent with randomness. Milojevic points to no effect of PM2.5 on heart attacks, RR= 1.003 with a p-value of 0.36.

Young et al. (2017) conducted a time series analysis of daily deaths for the eight most populated air basins in California. There is a total of over 37,000 exposure days. Three death classifications were analyzed, AllCause, Cardiovascular, and Respiratory. They found no effect of PM2.5 on acute deaths. Our p-value plot analysis supports the no effect claims of Milojevic and Young and call into question association or causation of PM with heart attacks.

In summary, the claims coming from the base papers used in the Nawrot meta-analysis are consistent with random results, those on the 45-degree line, and manipulated results, those on the blade of the hockey stick, while recent, large studies are consistent with no effect of PM2.5 on heart attacks. At this point, causality of PM10/PM2.5 on heart attacks is not supported.

**Contributors**
SSY designed the study. SSY, MKA and KD acquired, analysed, and interpreted data. SSY drafted the initial manuscript. SSY, MKA and KD edited the manuscript.

**Declaration of interests**
We declare no competing interests

**References, Lancet, Air quality and health effects, references numbers from Nawrot.**

**References, General**

Table 1. Authors, variable counts, and analysis search spaces for the 14 Nawrot base papers. The number before the author name is the Nawrot reference number.

| Author | Outcomes | Predictors | Lags | Covars | Space_1 | Space_2 | Space_3 |
|---|---|---|---|---|---|---|---|
| 12 Barnett | 7 | 4 | 1 | 13 | 28 | 8,192 | 229,376 |
| 13 Zanobetti | 2 | 6 | 2 | 7 | 24 | 128 | 3,072 |
| 14 Zanobetti | 5 | 19 | 1 | 5 | 95 | 32 | 3,040 |
| 15 Peters | 2 | 6 | 13 | 12 | 156 | 4,096 | 638,976 |
| 20 Lanki | 2 | 5 | 4 | 5 | 40 | 32 | 1,280 |
| 21 Cendon | 2 | 5 | 1 | 6 | 10 | 64 | 640 |
| 40 Koken | 5 | 6 | 10 | 5 | 300 | 32 | 9,600 |
| 41 Linn | 10 | 4 | 2 | 8 | 80 | 256 | 20,480 |
| 42 Mann | 21 | 4 | 7 | 9 | 588 | 512 | 301,056 |
| 43 Peters | 1 | 7 | 2 | 5 | 14 | 32 | 448 |
| 44 Pope | 1 | 2 | 7 | 9 | 14 | 512 | 7,168 |
| 45 Sullivan | 4 | 4 | 3 | 10 | 48 | 1,024 | 49,152 |
| 46 Ye | 8 | 5 | 5 | 5 | 200 | 32 | 6,400 |
| 47 Zanobetti | 5 | 2 | 4 | 7 | 40 | 128 | 5,120 |

Table 2 Summary statistics for the number of possible analyses using the three search spaces.

| Statistics | Space_1 | Space_2 | Space_3 |
|---|---|---|---|
| maximum | 588.0 | 8,192 | 638,976 |
| quartile | 167.0 | 640 | 94,208 |
| median | 44.0 | 128 | 6,784 |
| quartile | 21.5 | 32 | 2,600 |
| minimum | 10.0 | 32 | 448 |

Table 3. Named covariates used in studies.

| Paper | Covariates |
|---|---|
| 12 Barnett | Age, Sex, Region, T, T-1, RH, AP, DOW, HotDays, ColdDays, HD, HD-1, Rain |
| 13 Zanobetti | Age, Sex, T, AT, RH, DOW, season |
| 14 Zanobetti | T, DPT, Trend, Season, Year |
| 15 Peters | Age, Sex, T, Occupational status, Educational status, Smoking status, Hospital location, DOW, HOD, Season, Location, Symptoms of MI |
| 20 Lanki | Age, Sex, T, Season, Case fatality |
| 21 Cendon | mT, mT-1, RH, RH-1, DOW, Trend |
| 40 Koken | Age, Sex, T, DPT, DOW |
| 41 Linn | Age, Sex, T, RH, AP, Race, Region, Season |
| 42 Mann | Age, Sex, mT, RH, DOW, Year, Trend, Region, Day of Study |
| 43 Peters | Age, Sex, mT, RH, MH |
| 44 Pope | Age, Sex, T, DPT, Smoking status, BMI, Region, MH, Clearing index |
| 45 Sullivan | Age, Sex, T, RH, DOW, Race, BMI, Smoking status, MH, Season |
| 46 Ye | Age, Sex, MT, mT, Trend |
| 47 Zanobetti | Age, Sex, AT, AT-1, DOW, Season, Region |

03 Figures
Figure 1. P-value plot, ranked p-values plotted against the integers.

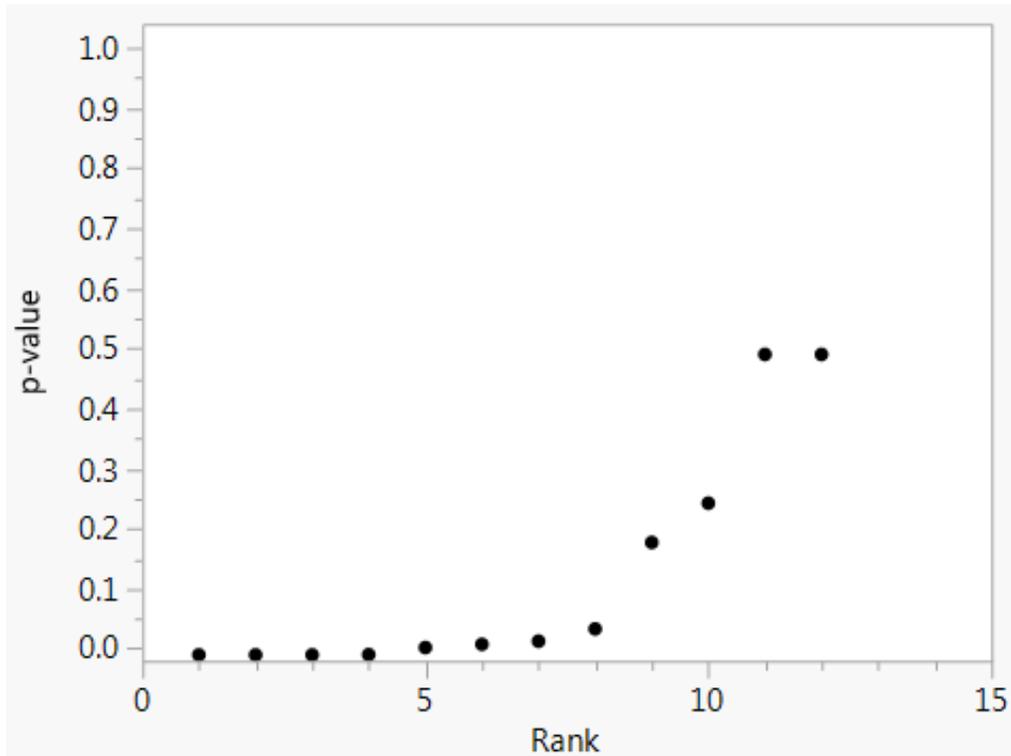

Note that two p-values were declared not significant in Nawrot, but p-values were not given so two not significant p-values are not plotted.

Figure 2.

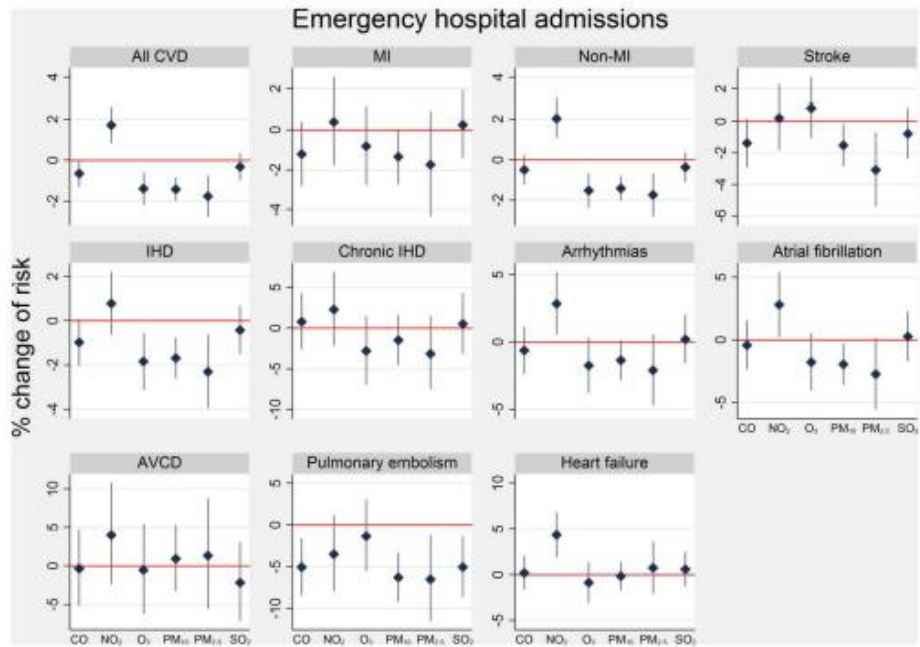

Figure 1  Per cent change (95% CI) in risk of emergency cardiovascular admissions for a 10th–90th centile change in pollutant at lags 0–4 days. 10th–90th centile ranges in pollutants for 2003–2008. AVCD, atrioventricular conduction disorder; MI, myocardial infarction; IHD, ischaemic heart disease. Data source: Hospital Episode Statistics database, 2003–2008.

Figure 3.

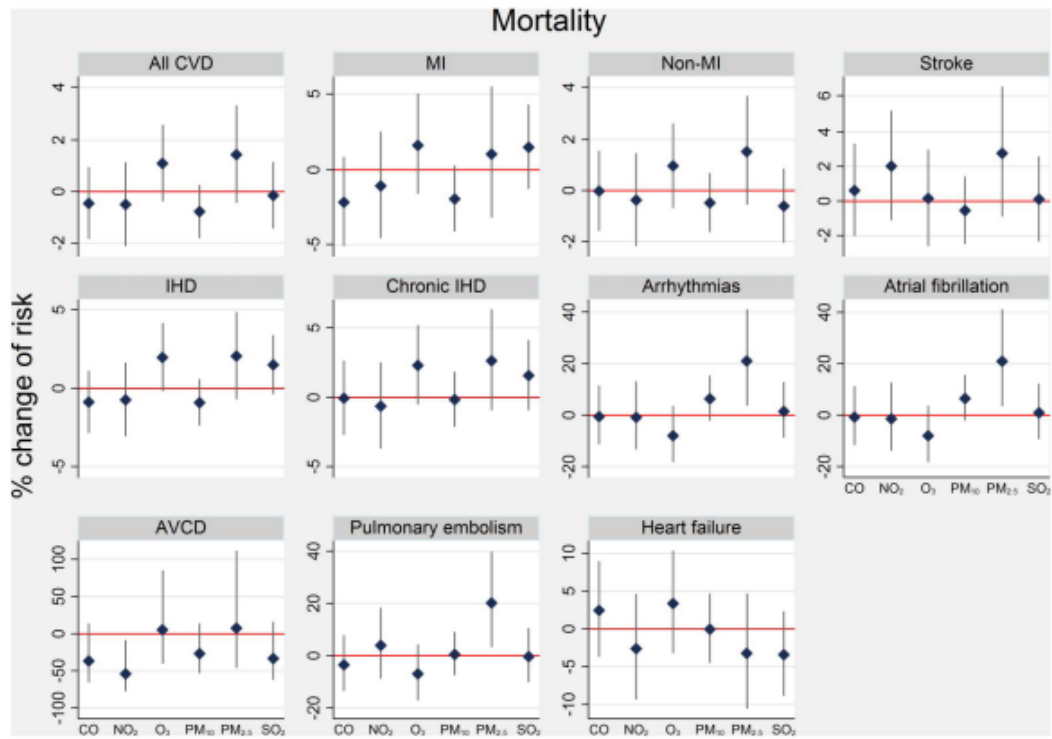

**Figure 2** Per cent change (95% CI) in risk of cardiovascular mortality for a 10th–90th centile change in pollutant at lags 0–4 days. 10th–90th centile ranges in pollutants for 2003–2006. AVCD, atrioventricular conduction disorder; MI, myocardial infarction; IHD, ischaemic heart disease. Data source: Office of National Statistics mortality registry, 2003–2006.